\documentclass[12pt,article,onecolumn]{IEEEtran}

\usepackage{srcltx,epsf,amsmath,amssymb,amsfonts,mathrsfs}
\usepackage{times}
\usepackage{subeqnarray}
\usepackage{epsfig}
\usepackage{url}
\usepackage{subfigure}
\usepackage{graphicx}
\usepackage{epstopdf}

\newtheorem{remark}{Remark}

\begin{document}

\title{Least-Squares Based Iterative Multipath Super-Resolution
Technique}
\author{\authorblockN{Wooseok Nam,~\IEEEmembership{Member,~IEEE,}
        Seung-Hyun~Kong*,~\IEEEmembership{Member,~IEEE}}\\
\authorblockA{Dept. of Aerospace Engineering, KAIST,\\ Daejeon, Republic of Korea\\
E-mail: nam.wooseok@gmail.com, skong@kaist.ac.kr}
\thanks{*Corresponding author.}} \maketitle

\begin{abstract}
In this paper, we study the problem of multipath channel
estimation for direct sequence spread spectrum signals. To resolve
multipath components arriving within a short interval, we propose
a new algorithm called the least-squares based iterative multipath
super-resolution (LIMS). Compared to conventional super-resolution
techniques, such as the multiple signal classification (MUSIC) and
the estimation of signal parameters via rotation invariance
techniques (ESPRIT), our algorithm has several appealing features.
In particular, even in critical situations where the conventional
super-resolution techniques are not very powerful due to limited
data or the correlation between path coefficients, the LIMS
algorithm can produce successful results. In addition, due to its
iterative nature, the LIMS algorithm is suitable for recursive
multipath tracking, whereas the conventional super-resolution
techniques may not be. Through numerical simulations, we show that
the LIMS algorithm can resolve the first arrival path among
closely arriving independently faded multipaths with a much lower
mean square error than can conventional early-late discriminator
based techniques.
\end{abstract}


\section{Introduction} \label{SEC:Introduction}

A pseudo-noise (PN) code sequence, which is a deterministic but
noise-like sequence generated by a linear feedback shift register,
is widely used in the areas of radar and sonar signal processing,
ranging and positioning, and digital communication, since it is
convenient to estimate the code phase of the received signal using
its peaky auto-correlation function (ACF). When the PN code
sequence is launched to a channel and received by a remote
terminal, the cross-correlation of the received signal and a
receiver replica of the PN code sequence yields a sharp peak at
the lag of the propagation delay caused by the channel. This delay
is related to the physical separation between the transmitter and
the receiver, and thus gives important information for source
localization or target detection. However, in many practical
situations, the propagation medium suffers from multiple echoes
with short delays due to scattering objects around the receiver.
In such a case, we are frequently interested in estimating the
delays of those echoes and, especially, the first arrival path,
since it is the most likely to be the line-of-sight path, which
contains the information of the true distance between the
transmitter and the receiver.

In the multipath channel, the conventional PN code phase
estimation method using the ACF, which finds the peak of the
correlator output, would probably give an incorrect estimate of
the first arrival path delay unless the first arrival path is the
most dominant path. This is because the ACF for the first arrival
path is buried under the sum of other overlapping ACFs of echos,
and thus the lag of the largest magnitude correlator output is
often different from the first arrival path delay. We can work
around this problem to some extent by finding multiple local
extrema of the correlator output and selecting the one with the
smallest lag as the estimate of the first arrival path
\cite{Rappaport89}. However, the time resolution of this method is
limited to the chip duration of the PN code sequence, which is not
sufficient in some narrow band applications.

In order to improve resolution of the conventional ACF based code
phase estimation, {\em super-resolution} techniques can be taken
into account. The super-resolution techniques, such as the
multiple signal classification (MUSIC) \cite{Schmidt86} and the
estimation of signal parameters via rotation invariance techniques
(ESPRIT) \cite{Roy89}, were originally proposed for direction
finding of multiple targets with passive sensor arrays, and first
applied to the channel multipath resolution in \cite{Bruck85}. In
\cite{Bruck85} and \cite{Kusuma03}, the super-resolution
techniques were applied directly to the received signal samples
without cross-correlation. In particular, in \cite{Kusuma03}, it
was shown that the super-resolution techniques can be applied to
received signals sampled at a considerably lower rate than the
chip rate of the PN code sequence when the signals have a finite
rate of innovation (FRI). On the other hand, in \cite{Manabe92,
Bouch01}, super-resolution techniques for the correlator output
signal were introduced. These techniques can significantly improve
the resolution of multipath delays, compared to conventional
techniques, but have some major drawbacks that render them
impractical. First, since the super-resolution techniques are
based on the subspace decomposition of the signal correlation
matrix, the signal components contributed by different channel
paths should occupy distinct dimensions of the subspace in order
to be distinguishable. For this, each distinct path should be far
enough apart from the others, and the attenuation of each path
should be varying randomly and not be perfectly correlated with
the others during an observation interval. These conditions are
not satisfied when the paths are quite narrowly spaced, the
channel is slowly varying, or the observation interval is not long
enough. In addition, it is not so reasonable to assume that only
the channel attenuations are randomly varying while the path
delays are fixed. As a result, additional techniques, such as
frequency smoothing \cite{Manabe92, Bouch01}, are required at the
transmitter to make the assumption valid. As a second drawback,
most super-resolution techniques are batch algorithms that compute
a set of estimates from a large number of stationary observations.
Although the technique for FRI signals \cite{Kusuma03} can work
with a very small number of observations, it has very low noise
immunity and also needs a large number of observations to achieve
robustness against noise. The radio ranging application, however,
often requires tracking of the signal as well. That is, a new
observation, which can be non-stationary, is periodically given,
and a new set of estimates should be immediately computed from the
new observation and the previous set of estimates in a recursive
manner \cite{Betz09p1, Betz09p2}. Thus, the super-resolution
techniques are not adequate for applications where tracking is a
concern.

In this paper, we consider a least-squares (LS) approach to
resolve the short-delay multipath components with a correlator
output signal. When the noise is Gaussian distributed, our
approach includes the maximum likelihood (ML) approach, which is
optimal in performance. The LS and ML approaches for the channel
multipath resolution were presented directly and indirectly in
earlier studies \cite{Kumar85, Bresler86, Stoica89}, but they have
not gathered much interest since the highly nonlinear nature in
the parameters to be estimated renders the use of these approaches
computationally intensive. In \cite{Kumar85} and \cite{Bresler86},
faster algorithms for solving a nonlinear LS problem were derived
by exploiting the linear predictability of sinusoidal signals, but
they are not applicable to our case. 
In our work, we propose an iterative algorithm to solve the LS
multipath resolution problem, which we will call the {\em
least-squares based iterative multipath super-resolution} (LIMS)
algorithm. Thanks to its iterative and recursive mode of
operation, it is appropriate for multipath tracking with
sequential observations of the received signal. Moreover, we show
that the tracking of a PN code phase using the early-late (EL)
discriminator \cite{Betz09p1, Betz09p2, Irsigler} is a special
case of our algorithm. Through numerical simulations, we show that
our scheme asymptotically achieves the optimal performance given
by the Cramer-Rao bound (CRB) at high carrier-to-noise density
ratio. In addition, we show that our algorithm outperforms the EL
discriminator based technique in the presence of severe
multipaths and noise. 

This paper is organized as follows. In Section \ref{SEC:SysModel},
we introduce the signal model and the LS formulation for the
multipath resolution. In Section \ref{SEC:SolvingLS}, we present
the LIMS algorithm as a technique to solve the LS multipath
resolution problem, and discuss the application of the LIMS
algorithm for multipath tracking. The complexity of the LIMS
algorithm and its relationship with the conventional EL
discriminator based technique are also discussed in Section
\ref{SEC:SolvingLS}. Results of numerical simulation in various
channel models and a comparison with the conventional method are
given in Section \ref{SEC:Simulation}, and Section
\ref{SEC:Conclusion} concludes the paper.

We will use the following notation throughout this paper. Vectors
or matrices are denoted by boldface symbols. The $p$-th element of
a vector and the $(p,q)$-th element of a matrix are denoted by
$[\cdot]_p$ and $[\cdot]_{p,q}$, respectively. The superscripts
$T$ and $H$ denote the transpose and conjugate transpose,
respectively. The Euclidean norm of a vector is denoted by $\|
\cdot \|$, and the infinity norm of a vector, i.e., the largest
absolute value of elements, is denoted by $\| \cdot \|_{\infty}$.
The fields of real and complex numbers are indicated by
$\mathbb{R}$ and $\mathbb{C}$, respectively, and the $N \times N$
identity matrix is denoted by ${\bf I}_N$. The statistical
expectation is denoted by $E\{\cdot\}$ and the real part of a
complex value is denoted by $\Re \{\cdot\}$. Finally, $O(\cdot)$
is the big O notation.


\section{System Model} \label{SEC:SysModel}

Consider a multipath channel between a transmitter and a receiver
with an impulse response
\begin{equation}
h(t) = \sum_{k=0}^{K-1} \gamma_k \delta(t - \tau_k ),\:\: 0 \leq t
< T_i \label{EQ:Channel} \text,
\end{equation}
where $K$ is the number of paths, $\left\{ \gamma_0, \ldots,
\gamma_{K-1} \right\}$ are the unknown complex channel
coefficients for each path, $\left\{ \tau_0, \ldots, \tau_{K-1}
\right\}$ are the unknown arrival time delays for each path,
$\delta (t)$ is the Dirac delta function, and $T_i$ ($\gg \max \{
\tau_0, \ldots, \tau_{K-1}\}$) is a time interval of interest. We
assume that the delays $\left\{ \tau_0, \ldots, \tau_{K-1}
\right\}$ are distinct and $\tau_0 < \tau_1 < \cdots <
\tau_{K-1}$, without loss of generality. Thus, the delay of the
first arrival path is $\tau_0$. The transmitter launches a
continuous time PN code sequence $s(t)$ with a chip duration $T_c$
to the channel (\ref{EQ:Channel}), and the baseband received
signal at the receiver is written as
\begin{align}
\eta (t) &= h(t) * s(t) + v(t) \nonumber\\
&= \sum_{k=0}^{K-1} \gamma_k s(t - \tau_k) + v(t),\:\: 0 \leq t <
T_i \text, \label{EQ:RxSig}
\end{align}
where $v(t)$ is a circularly symmetric, zero-mean, complex
Gaussian noise process with auto-correlation $N_0 \delta(\zeta)$.
For the first process at the receiver, the cross-correlation of
$\eta(t)$ with $s(t)$ is taken over the time interval $T_i$, and
the correlator output is uniformly sampled at $N$ lags
$\left\{0,T_s, \ldots, (N-1)T_s \right\}$ to yield
\begin{equation}
y_n = \sum_{k=0}^{K-1} \gamma_k R \left( nT_s - \tau_k
\right)+w_n,\:\: n = 0,\ldots,N-1 \text, \label{EQ:SigModel}
\end{equation}
where $R(\zeta)$ is the auto-correlation function (ACF) of the PN
code sequence given by
\begin{equation}
R(\zeta) = \frac{1}{T_i} \int_0^{T_i} s(t) s^* ( t - \zeta) d t
\text, \label{EQ:ACF1}
\end{equation}
and $w_n$ is the filtered noise process $w(\zeta)$ sampled at
$\zeta = nT_s$, where
\begin{equation}
w(\zeta) = \frac{1}{T_i} \int_0^{T_i} v(t) s^* (t - \zeta) d t
\text. \label{EQ:FilteredNoise}
\end{equation}
For a binary phase shift keying (BPSK) modulated PN code sequence
with a sufficiently large integration time $T_i \gg T_c$, the ACF
(\ref{EQ:ACF1}) is ideally given by
\begin{equation}
R(\zeta) = \begin{cases} \frac{\zeta}{T_c}+1, & -T_c < \zeta \leq
0 \text,\\
-\frac{\zeta}{T_c}+1, & 0< \zeta \leq T_c \text,\\
0, &\text{otherwise.} \end{cases}\label{EQ:AutoCorr}
\end{equation}
For notational simplicity in the subsequent expressions, we denote
${\bf y} = [y_0, \ldots, y_{N-1}]^T$, ${\bf c}_{\rm CH} =
[\gamma_0, \ldots, \gamma_{K-1}]^T$, ${\bf t}_{\rm CH} = [\tau_0,
\ldots, \tau_{K-1}]^T$, and ${\bf w} = [w_0, \ldots,
w_{N-1}]^T$. Then, from (\ref{EQ:FilteredNoise}), the covariance
matrix of the noise vector, $E \left\{ {\bf ww}^H \right\}$, is
given by $\frac{N_0}{T_i} {\bf C}$, where ${\bf C}$ has elements
\begin{equation}
\left[ {\bf C} \right]_{p,q} = \frac{T_i}{N_0}
E\{w_{p-1}w_{q-1}^*\} = R((p-q)T_s),\:\: 1 \leq p,q \leq N \text.
\label{EQ:NoiseCovMtx}
\end{equation}

From the above signal model, the LS estimation of the channel
coefficients and the path delays is formulated as
\begin{subequations}
\label{EQ:LSForm}
\begin{align}
\left\{ \hat{\bf c}, \hat{\bf t} \right\} &= \underset{{\bf
c},{\bf t}}{\arg\min} \left\| {\bf Gy}-{\bf GA} ({\bf t}) {\bf c}
\right\|^2 \label{EQ:LSForm1}\\
&= \underset{{\bf c},{\bf t}}{\arg\min} \left\| {\bf Gy}-{\bf G}
\left( {\bf B} ({\bf c},{\bf t}) {\bf t}+ {\bf b} ({\bf c}, {\bf
t}) \right) \right\|^2 \label{EQ:LSForm2} \text,
\end{align}
\end{subequations}
where ${\bf c} = [c_0,\ldots,c_{M-1}]^T \in \mathbb{C}^{M \times
1}$ and ${\bf t} = [t_0, \ldots, t_{M-1}]^T \in \mathbb{R}^{M
\times 1}$ are the vectors of channel coefficients and path delays
to be estimated, respectively, and $M$ is the assumed number of
multipaths for the estimation. Since the true number of
multipaths, $K$, is generally unknown, we assume that $M$ can be
different from $K$. Note that there are two equivalent LS
formulations (\ref{EQ:LSForm1}) and (\ref{EQ:LSForm2}), and the
related parameters are defined as
\begin{subequations}
\label{EQ:Param}
\begin{gather}
\left[ {\bf A} ({\bf t}) \right]_{n+1,m+1} = R \left( nT_s - t_m
\right) \text, \label{EQ:ParamA}\\
\left[ {\bf B} ({\bf c},{\bf t}) \right]_{n+1,m+1} = \begin{cases}
-\frac{c_m}{T_c}, &\frac{t_m-T_c}{T_s} < n \leq \frac{t_m}{T_s}
\text,\\
\frac{c_m}{T_c}, &\frac{t_m}{T_s} < n \leq \frac{t_m+T_c}{T_s}
\text,\\
0, & \text{otherwise,}
\end{cases} \label{EQ:ParamB}\\
{\bf b} ({\bf c},{\bf t}) = \sum_{m=0}^{M-1} {\bf b} (c_m, t_m)
\text, \label{EQ:Paramb}\\
\left[ {\bf b} (c_m, t_m) \right]_{n+1} = \begin{cases} c_m \cdot
\left( 1+\frac{nT_s}{T_c} \right), &\frac{t_m-T_c}{T_s} < n \leq
\frac{t_m}{T_s}
\text,\\
c_m \cdot \left( 1-\frac{nT_s}{T_c} \right), &\frac{t_m}{T_s} < n
\leq \frac{t_m+T_c}{T_s}
\text,\\
0, & \text{otherwise,}
\end{cases}\label{EQ:Parambb}\\
n = 0,\ldots,N-1, \;\; m = 0,\ldots,M-1 \nonumber \text,
\end{gather}
\end{subequations}
and ${\bf G} \in \mathbb{C}^{N \times N}$ is a proper nonsingular
weighting matrix. For the least-squares estimation
(\ref{EQ:LSForm}) to be feasible, we assume that $N \geq 2M$, and
${\bf A} ({\bf t})$ and ${\bf B} ({\bf c},{\bf t})$ have full
column rank. The first assumption can be reasonable by taking a
sufficient number of samples, and so is the second assumption when
the delays $\left\{ t_0, \ldots, t_{M-1} \right\}$ are distinct
and sufficiently spaced by a small fraction of $T_c$
\cite[2.4.7]{MunozText}.

\begin{remark}
The signal model (\ref{EQ:SigModel}) is restricted to using the
triangular shaped ACF (\ref{EQ:AutoCorr}) for simplicity, but it
can be extended to general ACFs through a proper linearization
around the sampling points. That is, for a general ACF $R(\zeta)$,
(\ref{EQ:ParamB}) and (\ref{EQ:Paramb}) can be replaced by
\begin{gather}
\left[ {\bf B} ({\bf c},{\bf t}) \right]_{n+1,m+1} = c_m
\cdot \frac{d R(nT_s - t_m) }{d t_m} \text,\\
\left[ {\bf b} ({\bf c},{\bf t}) \right]_{n+1} = \sum_{m=0}^{M-1}
c_m \cdot \left( R(nT_s - t_m) - \frac{d R(nT_s - t_m)
}{d t_m} t_m \right) \text.
\end{gather}
In addition, though we assume the uniform sampling with a sampling
period $T_s$ in (\ref{EQ:SigModel}), all the arguments in this
paper can be immediately extended to non-uniform sampling.
\end{remark}

\begin{remark}
If we let ${\bf G} = {\bf C}^{-\frac{1}{2}}$, the weighting matrix
${\bf G}$ whitens the noise and thus the LS formulation
(\ref{EQ:LSForm}) becomes the maximum likelihood (ML) formulation.
In Section \ref{SEC:Simulation}, we investigate the performance
with and without noise whitening by letting ${\bf G} = {\bf
C}^{-\frac{1}{2}}$ and ${\bf G} = {\bf I}_N$, respectively.
\label{REM:Whitening}
\end{remark}

\section{Solving the Least-Squares Multipath Resolution Problem}
\label{SEC:SolvingLS}

From the LS formulation (\ref{EQ:LSForm}), we can observe that it
is linear in ${\bf c}$ given ${\bf t}$ but nonlinear in ${\bf t}$
due to the dependency of ${\bf B}({\bf c},{\bf t})$ and ${\bf b}
({\bf c},{\bf t})$ on ${\bf t}$. Thus, to find the solution to the
LS problem, a computationally expensive multidimensional search
seems to be inevitable. However, there exists a useful workaround
to avoid an exhaustive search: the {\em gradient descent
algorithm} \cite{HaykinText}. The gradient descent algorithm can
efficiently find the local minimum of the LS problem in an
iterative manner, by stepping from the current guess of the LS
solution to the direction that decreases the LS weight in
(\ref{EQ:LSForm}) at every iteration. In the following subsection,
we propose a gradient descent algorithm for the LS multipath
resolution problem, which will be denoted by the least-squares
based iterative multipath super-resolution (LIMS) algorithm.

\subsection{Least-squares based iterative multipath super-resolution algorithm} \label{SEC:LIMR}

For notational simplicity, let $\tilde{\bf y} \triangleq {\bf
Gy}$, $\tilde{\bf A}({\bf t}) = {\bf GA}({\bf t})$, $\tilde{\bf
B}({\bf c},{\bf t}) = {\bf GB}({\bf c},{\bf t})$, and $\tilde{\bf
b}({\bf c},{\bf t}) = {\bf Gb}({\bf c},{\bf t})$ in
(\ref{EQ:LSForm}). The operation flow of the LIMS algorithm is
described as follows. First, at the $l$-th iteration ($l$ is a
positive integer), the intermediate guesses of the channel
coefficient and the path delay vectors, ${\bf c}_{l-1}$ and ${\bf
t}_{l-1}$, are given from the previous $(l-1)$-th iteration. Given
${\bf t}_{l-1}$, a new estimate of the channel coefficient vector,
${\bf c}_l$, is computed as
\begin{equation}
{\bf c}_l = \left( \tilde{\bf A}^H ({\bf t}_{l-1}) \tilde{\bf A}
({\bf t}_{l-1}) \right)^{-1} \tilde{\bf A}^H ({\bf t}_{l-1})
\tilde{\bf y} \text. \label{EQ:CIter1}
\end{equation}
Note that the matrix inversion in (\ref{EQ:CIter1}) exists with
assumptions that ${\bf G}$ is nonsingular and ${\bf A} ({\bf t})$
has full column rank. The new estimate ${\bf c}_l$
(\ref{EQ:CIter1}) minimizes the LS weight in (\ref{EQ:LSForm1})
for a given ${\bf t}_{l-1}$, but it requires a matrix inversion
that can sometimes be troublesome in terms of complexity or
numerical stability issues. Thus, as an alternative, a refinement
by the gradient descent algorithm can be used as
\begin{equation}
{\bf c}_l = {\bf c}_{l-1} + \alpha \cdot \tilde{\bf A}^H ({\bf
t}_{l-1}) \left( \tilde{\bf y} - \tilde{\bf A} ({\bf t}_{l-1})
{\bf c}_{l-1} \right) \text, \label{EQ:CIter2}
\end{equation}
where $\alpha$ is a properly chosen step size. The refinement
${\bf c}_l$ (\ref{EQ:CIter2}) does not have complexity or
stability problems like (\ref{EQ:CIter1}), but may converge more
slowly than (\ref{EQ:CIter1}).

For the second step, a refinement ${\bf t}_l$ is computed from
${\bf t}_{l-1}$ and ${\bf c}_l$ using the gradient descent
algorithm, for which it is required to find the gradient of the LS
weight in (\ref{EQ:LSForm2}) with respect to ${\bf t}$. However,
since the LS weight is nonlinear and non-analytic in ${\bf t}$, it
is not easy to compute the exact gradient.  Thus, we compute an
approximate gradient on the basis of the following assumptions:
For a small perturbation vector ${\bf e}$,
\begin{subequations}
\label{EQ:LSAssum}
\begin{align}
{\bf B}({\bf c},{\bf t}+{\bf e}) \simeq {\bf B}({\bf c},{\bf
t})\text,
\label{EQ:LSAssum1}\\
{\bf b}({\bf c},{\bf t}+{\bf e}) \simeq {\bf b}({\bf c},{\bf t})
\text. \label{EQ:LSAssume2}
\end{align}
\end{subequations}
From (\ref{EQ:ParamB}), (\ref{EQ:Paramb}), and (\ref{EQ:Parambb}),
the assumptions in (\ref{EQ:LSAssum}) are quite reasonable for a
magnitude of ${\bf e}$ sufficiently small in the sense that $\|
{\bf e} \|_{\infty} \ll T_s$. Under these assumptions, it can be
assumed that $\tilde{\bf B} ({\bf c}_l,{\bf t}_{l-1})$ and
$\tilde{\bf b} ({\bf c}_l,{\bf t}_{l-1})$ are constant in a small
region around ${\bf t}_{l-1}$. As a result, the gradient descent
for ${\bf t}$ reduces to
\begin{align}
{\bf t}_l &= {\bf t}_{l-1} + \beta \cdot \Re \left\{ \tilde{\bf
B}^H ({\bf c}_l,{\bf t}_{l-1}) \left( \tilde{\bf y} - \left(
\tilde{\bf B} ({\bf c}_l,{\bf t}_{l-1}) {\bf t}_{l-1} + \tilde{\bf
b} ({\bf
c}_l,{\bf t}_{l-1}) \right) \right) \right\} \nonumber\\
&= {\bf t}_{l-1} + \beta \cdot \Re \left\{ \tilde{\bf B}^H ({\bf
c}_l,{\bf t}_{l-1}) \left( \tilde{\bf y} - \tilde{\bf A} ({\bf
t}_{l-1}) {\bf c}_l \right) \right\} \text, \label{EQ:TIter}
\end{align}
where $\beta$ is a proper step size. The step size $\beta$ should
be small enough to satisfy the assumptions in (\ref{EQ:LSAssum}),
but a systematic criterion for $\beta$ that guarantees the
convergence is not easy to find.

Upon getting ${\bf c}_l$ using (\ref{EQ:CIter1}) or
(\ref{EQ:CIter2}), and ${\bf t}_l$ using (\ref{EQ:TIter}), the
iteration index $l$ is increased by one and the same computations
are repeated for ${\bf c}_{l+1}$ and ${\bf t}_{l+1}$. In this way,
the iteration continues until a specific stop criterion is met.
There are a number of stop criteria for this kind of iterative
algorithm, e.g., reaching a fixed number of iterations,
thresholding the magnitudes of changes of variables, etc., and we
omit a detailed discussion of these criteria. The LIMS algorithm
is summarized in Table \ref{TAB:LIMR}.

\begin{table} [b]
\begin{center}
\caption{Summary of the LIMS algorithm.} \label{TAB:LIMR}
\begin{tabular}{l}
\hline {\em Input}\\
\hspace{5mm}${\bf y} = [y_0, \ldots, y_{N-1}]^T$\vspace{1mm}\\
{\em Known parameters}\\
\hspace{5mm}Weighting matrix: ${\bf G} \in \mathbb{C}^{N \times
N}$\\
\hspace{5mm}Step sizes: $\alpha$ and $\beta$\vspace{1mm}\\
{\em Initial values}\\
\hspace{5mm}${\bf c}_0$ and ${\bf t}_0$\vspace{1mm}\\
{\em Computation}: $l=1,2,\ldots$ \\
\hspace{5mm}${\bf c}_l = \left( \tilde{\bf A}^H ({\bf t}_{l-1})
\tilde{\bf A} ({\bf t}_{l-1}) \right)^{-1} \tilde{\bf A}^H ({\bf
t}_{l-1}) \tilde{\bf y}$ \\
\hspace{5mm}$\left( \text{or}\:\: {\bf c}_l = {\bf c}_{l-1} +
\alpha \cdot \tilde{\bf A}^H ({\bf t}_{l-1}) \left( \tilde{\bf y}
- \tilde{\bf A} ({\bf t}_{l-1}) {\bf c}_{l-1} \right) \right)$\\
\hspace{5mm}${\bf t}_l = {\bf t}_{l-1} + \beta \cdot \Re \left\{
\tilde{\bf B}^H ({\bf c}_l,{\bf t}_{l-1}) \left( \tilde{\bf y} -
\tilde{\bf A} ({\bf t}_{l-1}) {\bf c}_l \right) \right\}$\vspace{1mm}\\
{\em Definitions}\\
\hspace{5mm}${\bf A} ({\bf t})$, ${\bf B} ({\bf c},{\bf t})$: given in (\ref{EQ:Param})\\
\hspace{5mm}$\tilde{\bf y} \triangleq {\bf Gy}$, $\tilde{\bf
A}({\bf t}) = {\bf GA}({\bf t})$, $\tilde{\bf B}({\bf c},{\bf t})
= {\bf GB}({\bf
c},{\bf t})$\\
 \hline
\end{tabular}
\end{center}
\end{table}


\subsection{Complexity issues}

The major burden of the LIMS algorithm is the matrix inversion in
(\ref{EQ:CIter1}), which is $O(M^3)$ of computation per iteration.
The true number of multipaths, $K$, could be large in practice
and, hence, the assumed number of multipaths, $M$, should be
accordingly large. However, assuming that only a few of the
multipaths are dominant, we can keep $M$ small; say, $M<10$.
Therefore, when $N \gg M$, the matrix inversion may not be too
burdensome compared to the overall complexity. In addition, as
noted in Section \ref{SEC:LIMR}, the matrix inversion can be
avoided by using (\ref{EQ:CIter2}) instead, which is $O(NM)$ of
computation per iteration. The total number of iterations is
another important factor for the complexity. However, as will be
discussed in Section \ref{SEC:Tracking}, the number of iterations
can be considerably smaller when the LIMS algorithm is combined
with a tracking function such as the delay locked loop.


\subsection{Multipath tracking} \label{SEC:Tracking}

In many practical realizations of radio ranging systems, a new
observation of the received signal is obtained periodically.
Successive observations may include different multipath channels
when the channel is time varying. Under the assumption that the
channels for the successive observations are not statistically
independent, the time varying path delays can be efficiently
tracked. One such tracking system is the delay locked loop (DLL)
\cite{Betz09p1, Betz09p2}. In the conventional DLL used for PN
code phase tracking, the early-late (EL) discriminator is widely
used to generate an error signal from the previous estimate of the
path delay and the new observation of the received signal
(correlator output). The error signal is then filtered by the loop
filter in the tracking loop and a refined delay estimate is
computed. Now, instead of the EL discriminator, we consider using
the LIMS algorithm in the tracking loop.

Suppose that the $\nu$-th observation of the received signal
samples, ${\bf y}^{(\nu)}$, is given. Using the LIMS algorithm,
estimates of the channel coefficient vector $\hat{\bf c}^{(\nu)}$
and the path delay vector $\hat{\bf t}^{(\nu)}$ are computed.
Then, when the $(\nu+1)$-th observation ${\bf y}^{(\nu +1)}$ is
given, the LIMS algorithm starts a new round of iterations with
$\hat{\bf c}^{(\nu)}$ and $\hat{\bf t}^{(\nu)}$ as initial values.
As long as $\hat{\bf c}^{(\nu)}$ and $\hat{\bf t}^{(\nu)}$ are
good estimates of the channel coefficients and the path delays,
and the channels for the $\nu$-th and the $(\nu+1)$-th
observations do not differ much, the new estimates $\hat{\bf
c}^{(\nu+1)}$ and $\hat{\bf t}^{(\nu+1)}$ might be obtained in a
small number of iterations.


When the channel is not time-varying but corrupted by different
realizations of noise at different observations, applying the LIMS
algorithm with a small number of iterations for each observation
can be seen as the application of the {\em stochastic gradient
method} for the channel estimation \cite{HaykinText}. Thus, as the
number of observations increases, the accuracy of channel
estimation can improve until it reaches the steady state.

\subsection{Relationship with the conventional early-late
discriminator} \label{SEC:GeneralDisc}

\psfull
\begin{figure} [t]
\begin{center}
\epsfig{file=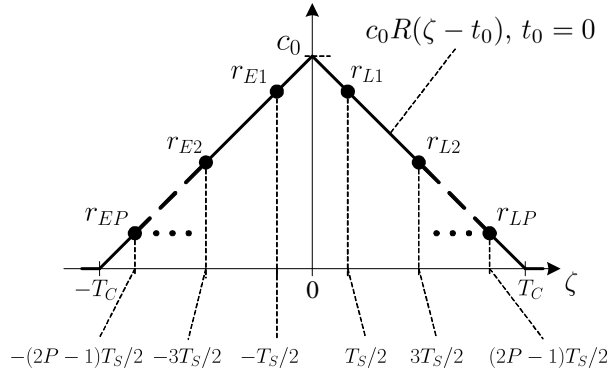, width=8cm}
\caption{Sampling $c_0 R(\zeta-t_0)$ at $2P$ points.}
\label{FIG:CorrTrack}
\end{center}
\end{figure}
\psdraft

The EL discriminator computes the difference between the powers
(magnitude squares) of two samples with different lags at the
correlator output. The two samples are called the early and the
late samples and are separated by a constant interval. When the
peak of an ACF is located in between the two samples, the
difference between the two powers serves as a feedback signal that
shifts the sampling points so that the peak of the ACF is at the
midpoint of the sampling points. Intuitively, this operation is
established on the assumptions that the channel has a single path
and that the correlator output maintains the exact triangular
shape of the ACF $R(\zeta)$ (\ref{EQ:AutoCorr}). However, these
assumptions are not valid when the channel has multiple paths and
the received signal is corrupted by noise.

Basically, the EL discriminator can be thought of as a special
case of the LIMS algorithm with the single path assumption, i.e.,
$M=1$, and two samples, i.e., $N=2$. To illustrate this
relationship, let us first consider $2P$ sampling points at
$\zeta=\frac{(2p+1)T_s}{2}$, $p=-P,\ldots,P-1$, where $P =
\left\lfloor \frac{T_c}{T_s}+\frac{1}{2} \right\rfloor$. Then, at
the sampling points, the sampled values of the correlator output
are denoted by $y_{E(-p)}$ for $p < 0$ and $y_{L(p+1)}$ for $p
\geq 0$. Likewise, the sampled values of the scaled ACF, $c_0
R(\zeta-t_0)$, are denoted by $r_{E(-p)}$ for $p < 0$ and
$r_{L(p+1)}$ for $p \geq 0$, as shown in Fig. \ref{FIG:CorrTrack}.
Now, we temporarily let $T_s = T_c$, and thus $P=1$ and the number
of total samples is $N=2P=2$. Assuming $-\frac{T_c}{2} < t_0 \leq
\frac{T_c}{2}$ and the number of paths $M=1$, the LS weight is
given by
\begin{subequations}
\label{EQ:LSforEL}
\begin{align}
&\left\| \begin{bmatrix} y_{E1} \\ y_{L1}
\end{bmatrix} -
\begin{bmatrix} r_{E1} \\ r_{L1} \end{bmatrix} \right\|^2 \\
=& \left\| \begin{bmatrix} y_{E1} \\ y_{L1} \end{bmatrix} -
c_0 \begin{bmatrix} 0.5 - t_0 \\ 0.5 + t_0 \end{bmatrix} \right\|^2 \\
=& \left\| \begin{bmatrix} y_{E1} \\ y_{L1} \end{bmatrix} - c_0
\begin{bmatrix} -1 \\ 1 \end{bmatrix} t_0 - c_0 \begin{bmatrix} 0.5
\\ 0.5 \end{bmatrix} \right\|^2 \text.
\end{align}
\end{subequations}
Note that, since the two samples are one chip apart, the noise
levels for the two samples are independent according to
(\ref{EQ:NoiseCovMtx}). Hence, by Remark \ref{REM:Whitening},
(\ref{EQ:LSforEL}) is the ML weight as well. From
(\ref{EQ:Param}), parameter vectors for the LS weight
(\ref{EQ:LSforEL}) are defined as
\begin{subequations}
\label{EQ:ELParam}
\begin{align}
{\bf A}(t_0) &= \begin{bmatrix} 0.5 - t_0 \\ 0.5 + t_0
\end{bmatrix} \text, \\
{\bf B} (c_0,t_0) &= c_0 \begin{bmatrix} -1 \\ 1 \end{bmatrix} \text,\\
{\bf b} (c_0,t_0) &= c_0 \begin{bmatrix} 0.5 \\ 0.5
\end{bmatrix} \text.
\end{align}
\end{subequations}
Applying the LIMS algorithm to the LS weight (\ref{EQ:LSforEL})
with the initial value $t_{0,0} = 0$, the first iteration gives
\begin{equation}
c_{0,1} = \left( {\bf A}^H (t_{0,0}) {\bf A} (t_{0,0})
\right)^{-1} {\bf A}^H (t_{0,0}) \begin{bmatrix} y_{E1} \\ y_{L1}
\end{bmatrix} = y_{E1} + y_{L1}
\end{equation}
by (\ref{EQ:CIter1}), and $t_{0,1} = t_{0,0} - \beta D$ by
(\ref{EQ:TIter}), where
\begin{align}
D &= -\Re \left\{ {\bf B}^H (c_{0,1},t_{0,0}) \left( \begin{bmatrix} y_{E1} \\
y_{L1} \end{bmatrix} - c_{0,1} {\bf A} (t_{0,0}) \right) \right\} \nonumber\\
&= -\Re \left\{ \left( y_{E1} + y_{L1} \right)^* \left( y_{E1} - y_{L1} \right) \right\} \nonumber\\
&= \left| y_{E1} \right|^2 - \left| y_{L1} \right|^2 \text.
\label{EQ:ELCorr}
\end{align}
As expected, $D$ (\ref{EQ:ELCorr}), the approximate gradient with
respect to $t_0$, is the same as the EL discriminator.

\psfull
\begin{figure} [t]
\begin{center}
\epsfig{file=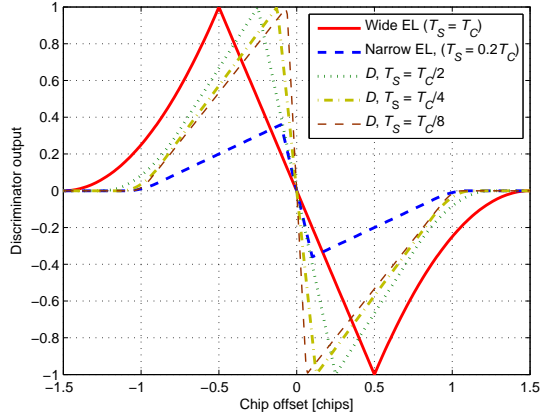, width=8cm}
\caption{Discriminator responses.} \label{FIG:ELCorr}
\end{center}
\end{figure}
\psdraft

Similarly, for $T_s < T_c$, we can derive
\begin{equation}
D = -\Re \left\{ {\bf B}^H (c_{0,1},0) {\bf C}^{-1} \left( {\bf y}
- c_{0,1} {\bf A} (0) \right) \right\} \text,
\label{EQ:GeneralDisc1}
\end{equation}
where ${\bf y} = [y_{EP},\ldots,y_{E1},y_{L1},\ldots,y_{LP} ]^T$,
\begin{equation}
c_{0,1} = \left( {\bf A}^H (0) {\bf C}^{-1} {\bf A} (0)
\right)^{-1} {\bf A}^H (0) {\bf C}^{-1} {\bf y} \text,
\label{EQ:GeneralDisc2}
\end{equation}
and ${\bf A}(0)$ and ${\bf B} (c_{0,1},0)$ are defined similarly
to (\ref{EQ:ELParam}), by letting $M=1$ in (\ref{EQ:Param}). Note
that, in (\ref{EQ:GeneralDisc1}) and (\ref{EQ:GeneralDisc2}),
noise whitening is applied. Then, (\ref{EQ:GeneralDisc1}) can be
regarded as a generalized $2P$-point discriminator as an extension
of the conventional $2$-point discriminator like the EL
discriminator. The quantity $D$ as a function of the code phase
offset for different values of $T_s$ is evaluated and plotted in
Fig. \ref{FIG:ELCorr}. The discriminator responses of the EL
discriminators with sample spacing $T_c$ (wide discriminator) and
$0.2T_c$ (narrow discriminator) are also plotted in Fig.
\ref{FIG:ELCorr} for comparison. In the figure, it is observed
that, as $T_s$ decreases, the discriminator curve gets steeper
around the zero offset, which implies stronger resistance against
perturbation after the DLL is locked at the peak of the ACF
\cite{Irsigler}.

\section{Numerical results} \label{SEC:Simulation}

This section presents simulation results to demonstrate the
performance of the LIMS algorithm. For comparison, the
conventional EL discriminators for PN code phase tracking with
wide ($T_c$) and narrow ($0.2 T_c$) spacings are also considered. In
particular, performance is evaluated with the estimation accuracy
of the delay of the first arrival path and, as the performance
measure, the mean square error (MSE) of the delay estimate is
used. For all simulations, the results are obtained by averaging
over $10^4$ trials with independent realizations of noise and
channel impulse response. As for the PN code sequence, we use the
global positioning system (GPS) coarse-acquisition (C/A) code
sequence \cite{MisraText}, which has 1023 chips per period with a
$1.023$MHz chip rate; one chip occupies $T_c =
1/(1.023\times10^6)$ seconds. The received signal is generated by
superimposing the attenuated, phase rotated, and delayed replicas
of the C/A code sequence according to the multipath channel
realization, and adding white Gaussian noise with the power
spectral density $N_0$. The received carrier power is defined as
$C = E \left\{ |h(t) * s(t) |^2 \right\}$ in (\ref{EQ:RxSig}),
where the expectation is over the channel realizations. The carrier
power $C$ varies resulting in a carrier to
noise density ratio ($C/N_0$) between $20$dBHz and $60$dBHz. At
the receiver, the integration time of the correlator, $T_i$, is
set to be $10$ms, which corresponds to $10$ periods
of the GPS C/A code sequence. 

For all simulations, it is assumed that a window of the correlator
output signal of $3T_c$ width is available for the EL
discriminator and the LIMS algorithm. The center of the window is
defined as the zero lag, i.e., $\zeta = 0$. It is also assumed
that the coarse acquisition of the PN code phase is perfect
regardless of $C/N_0$ so that the true first arrival path is
placed at a random lag uniformly distributed over $[-0.5T_c,
0.5T_c)$. Note that, since the size of the observation is too
small to compute the sample correlation matrix of the received
signal, conventional super-resolution techniques like MUSIC and
ESPRIT cannot be applied to the considered simulation environment.



\subsection{Simulation system}

For the simulation of the LIMS algorithm, the sampling period $T_s
= 0.1T_c$ is used. As described in Section \ref{SEC:LIMR},
equation (\ref{EQ:CIter1}) is used for the channel coefficient
update and, for ease of implementation, the signed gradient
descent algorithm is used in place of (\ref{EQ:TIter}) for the
delay update. That is,
\begin{equation}
{\bf t}_l = {\bf t}_{l-1} + \beta \cdot {\rm sgn} \left[ \Re
\left\{ \tilde{\bf B}^H ({\bf c}_l,{\bf t}_{l-1}) \left(
\tilde{\bf y} - \tilde{\bf A} ({\bf t}_{l-1}) {\bf c}_l \right)
\right\} \right]\text, \label{EQ:SignedTIter}
\end{equation}
where ${\rm sgn}[\cdot]$ is the element-wise sign function. For
the step size $\beta$ in (\ref{EQ:SignedTIter}), we use $\beta =
T_s/64$. To start the iteration, the initial value ${\bf t}_0 =
[t_{0,0},\ldots,t_{M-1,0}]^T$ is given such that $t_{0,0} =
-0.5T_c$, $t_{M-1,0} = 0.5T_c$, and $t_{m,0}$, $0<m<M-1$ evenly
divide the interval $(-0.5T_c,0.5T_c)$. With this initial value,
the final estimates of the channel coefficient vector $\hat{\bf c}
= [\hat{c}_0, \ldots, \hat{c}_{M-1}]^T$ and the delays vector
$\hat{\bf t} = [\hat{t}_0, \ldots, \hat{t}_{M-1}]^T$ are obtained
by ${\bf c}_{500}$ and ${\bf t}_{500}$, respectively.

Upon completion of the estimation process, a simple path
validation process follows to reduce the probability of false
alarm; an estimated path is valid if the estimated power ($|
\hat{c}_m |^2$) of the path is larger than $1$\% of the total
estimated power ($\sum_{m=0}^{M-1} | \hat{c}_m |^2$). The path
with the smallest delay among the valid paths is chosen for the
first arrival path.

Finally, in the simulations of the LIMS algorithm, two different
cases, without noise whitening, i.e., ${\bf G} = {\bf I}_M$, and
with noise whitening, i.e., ${\bf G} = {\bf C}^{-1/2}$, are
included for comparison.

\subsection{Two-path non-fading channel} \label{SEC:SimNonFade}

The first test channel considered for simulation is a two-path
non-fading channel with ${\bf c}_{\rm CH} = \left[
\frac{1}{\sqrt{2}}, \frac{1}{\sqrt{2}} \right]^T$ and ${\bf
t}_{\rm CH} = \left[ 0, \frac{T_c}{2} \right]^T$. The performance
results in terms of the MSE with respect to $C/N_0$ are shown in
Fig. \ref{FIG:TestCh1A}. As a benchmark, the CRB, which is derived
in Appendix
, is also plotted in the same figure. With this channel, the
correlator output has a flat top when there is no noise. As a
result, the conventional EL discriminator, which assumes a single
path and the triangular ACF, cannot perform well with the channel.
On the other hand, the LIMS algorithm can handle the two paths
simultaneously and outperforms the EL discriminator. It is also
observed that the LIMS algorithm with noise whitening consistently
performs better than that without noise whitening. In particular,
at high $C/N_0$'s, the LIMS algorithm with noise whitening
asymptotically achieves the CRB.


In Fig. \ref{FIG:TestCh1B}, learning curves of the LIMS algorithm
with noise whitening for the first arrival path delay are depicted
for $50$ independent trials at $C/N_0 = 30$dBHz. Observe that,
even with a very large initial deviation, the algorithm
successfully converges to the vicinity of the true first arrival
path delay within $300$ iterations.

\psfull
\begin{figure} [t]
\begin{center}
\subfigure[Mean square errors]{\epsfig{file=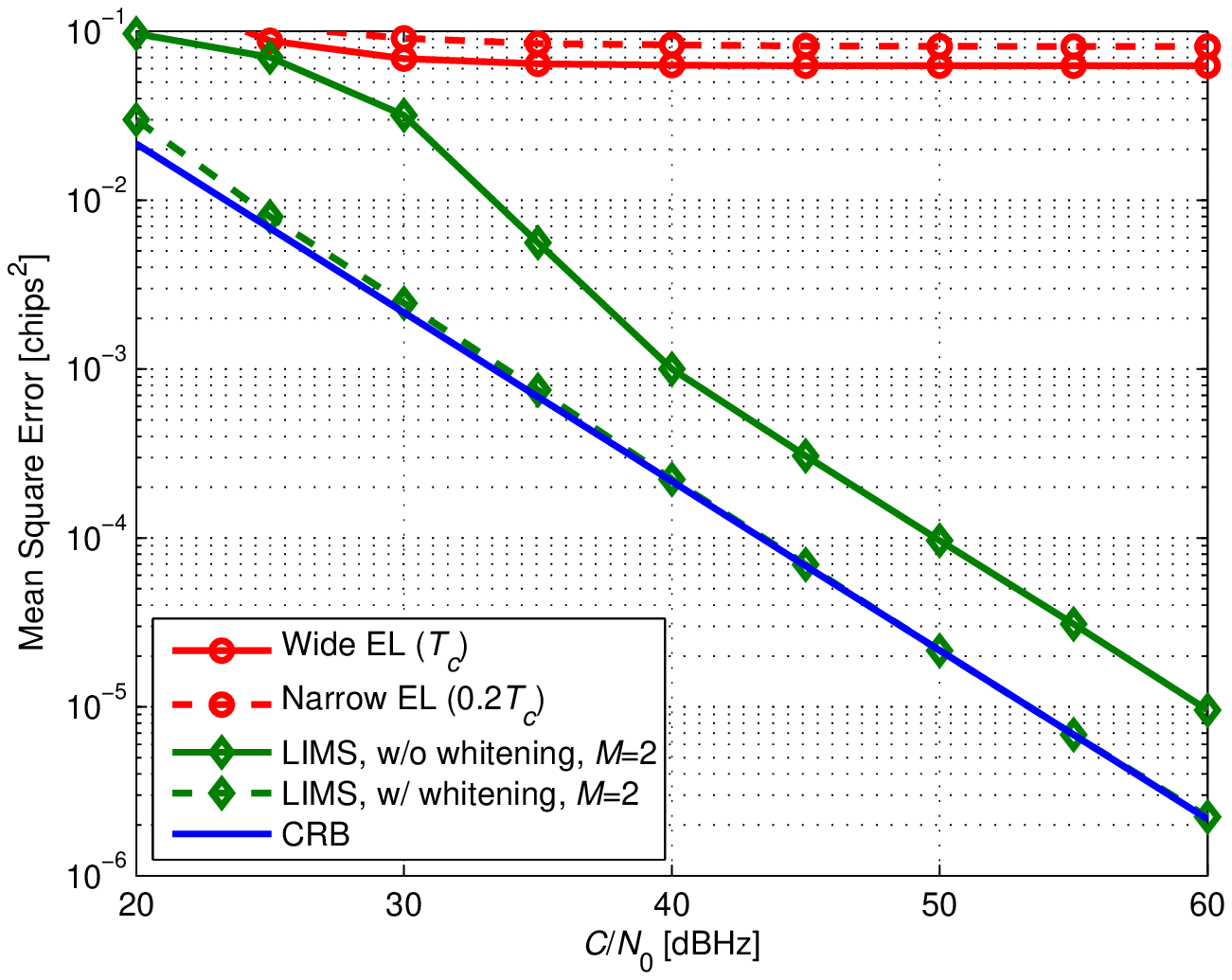,
width=8cm} \label{FIG:TestCh1A}} \subfigure[Learning curves at
$C/N_0=30$dBHz]{\epsfig{file=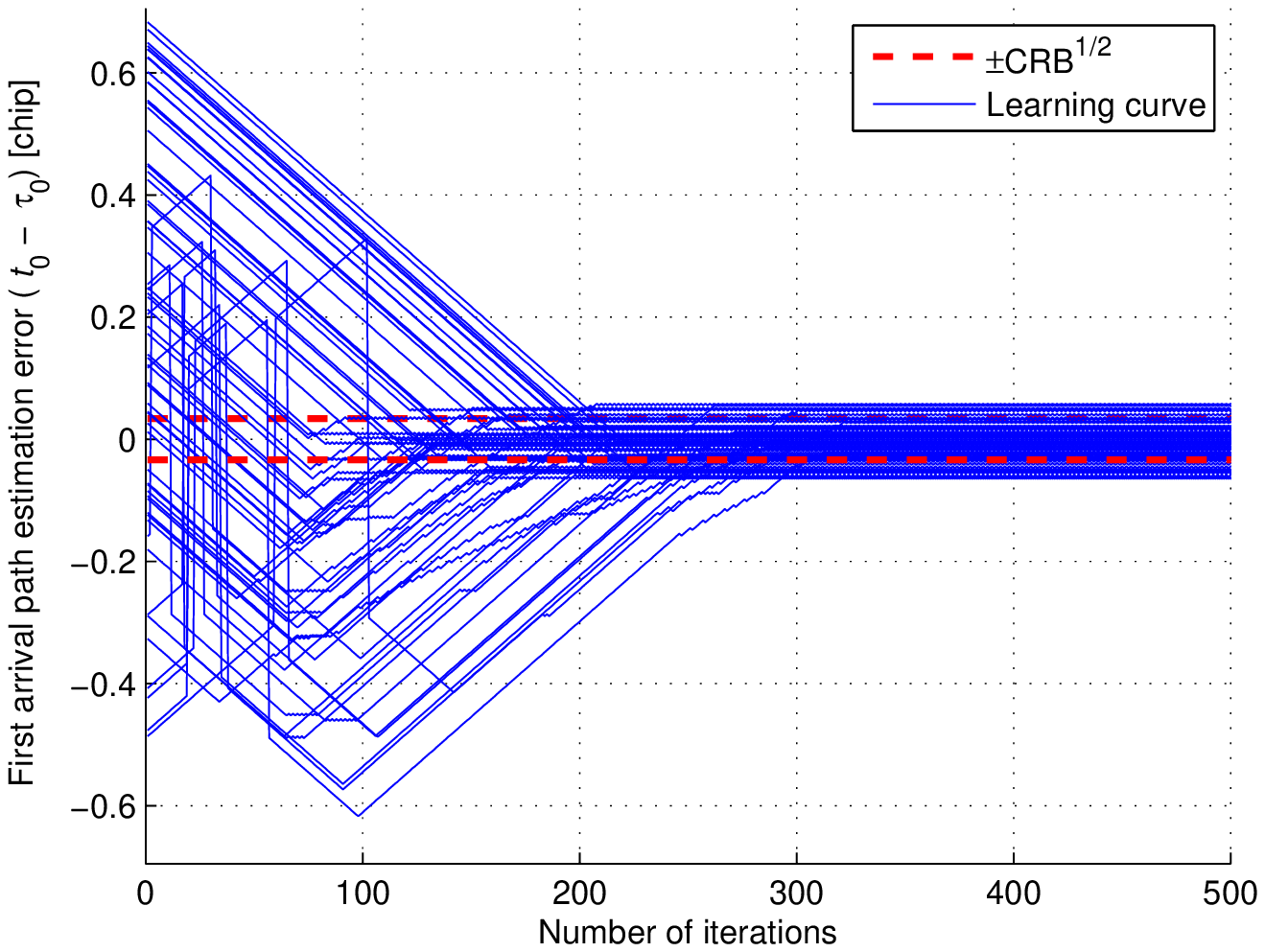, width=8cm}
\label{FIG:TestCh1B}} \caption{Mean square errors and learning
curves of the first arrival path delay estimation: ${\bf c}_{\rm
CH} = \left[ 1/\sqrt{2}, 1/\sqrt{2} \right]^T$ and ${\bf t}_{\rm
CH} = \left[ 0, {T_c}/{2} \right]^T$.} \label{FIG:TestCh1}
\end{center}
\end{figure}
\psdraft

\subsection{Single-path fading channel}

\begin{figure} [t]
\begin{center}
\epsfig{file=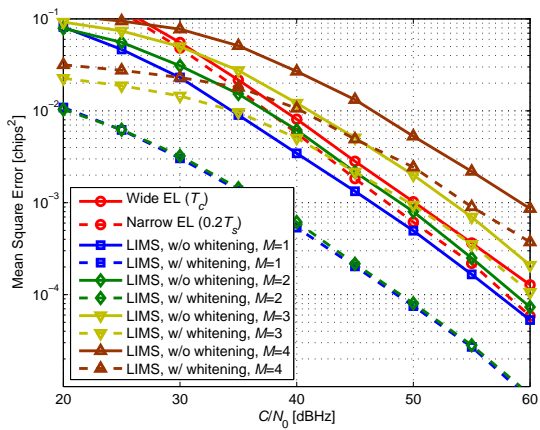, width=8cm} \caption{Mean square
errors of the first arrival path delay estimation: Single-path
fading channel.} \label{FIG:SinglePath}
\end{center}
\end{figure}

For the second simulation, a channel with only one path is
considered, where the channel coefficient $\gamma_0$ is a
zero-mean circular symmetric complex Gaussian random variable with
$E \left\{ | \gamma_0 |^2 \right\}=1$. The MSE performances of the
EL discriminators and the LIMS algorithms with different settings
for this channel are plotted in Fig. \ref{FIG:SinglePath}. As
shown in the figure, the EL discriminators, especially the one
with narrow spacing, perform very well for the channel. This is
because the assumption that the correlator output has a triangular
shape, on which the EL discriminator is based, is valid for this
case. Among all performance curves plotted in Fig.
\ref{FIG:SinglePath}, the best is achieved by the LIMS algorithm
with $M=1$ and noise whitening. Compared to the performance of the
EL discriminators, the best performance is an order of magnitude
superior. In terms of the generalized discriminator presented in
Section \ref{SEC:GeneralDisc}, this performance improvement arises
from the fact that the LIMS algorithm can make use of a larger
number of received signal samples and thus has a steeper
discriminator response. Finally, in Fig. \ref{FIG:SinglePath},
note that the LIMS algorithm with $M=2$ and noise whitening has a
performance comparable to that with $M=1$ and noise whitening even
though the number of assumed paths is not the same as the true
value.


\psfull
\psdraft

\subsection{Multipath fading channels}

\begin{figure} [t]
\begin{center}
\subfigure[Channel A]{\epsfig{file=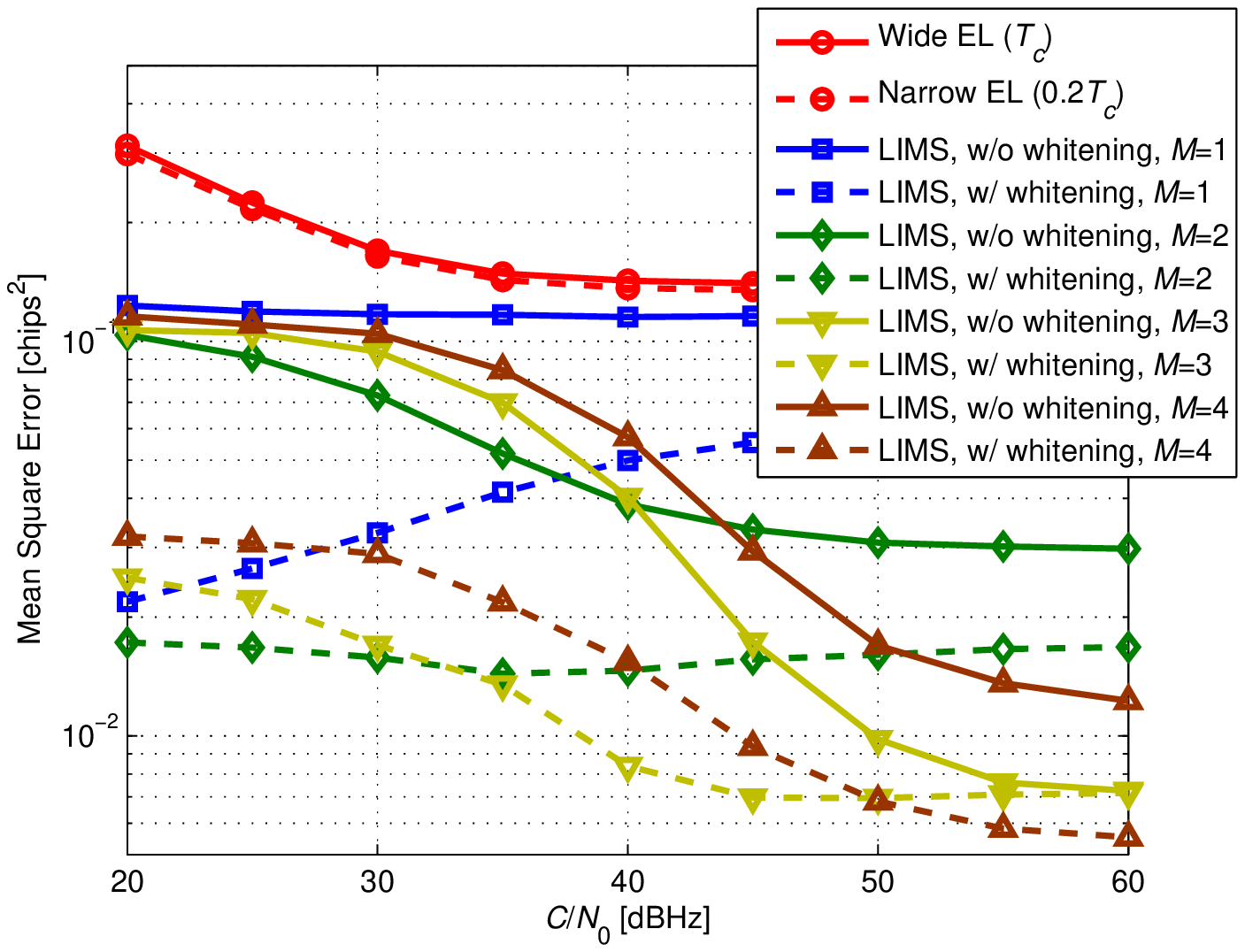,
width=8cm}\label{FIG:MultipathChInfA}} \subfigure[Channel
B]{\epsfig{file=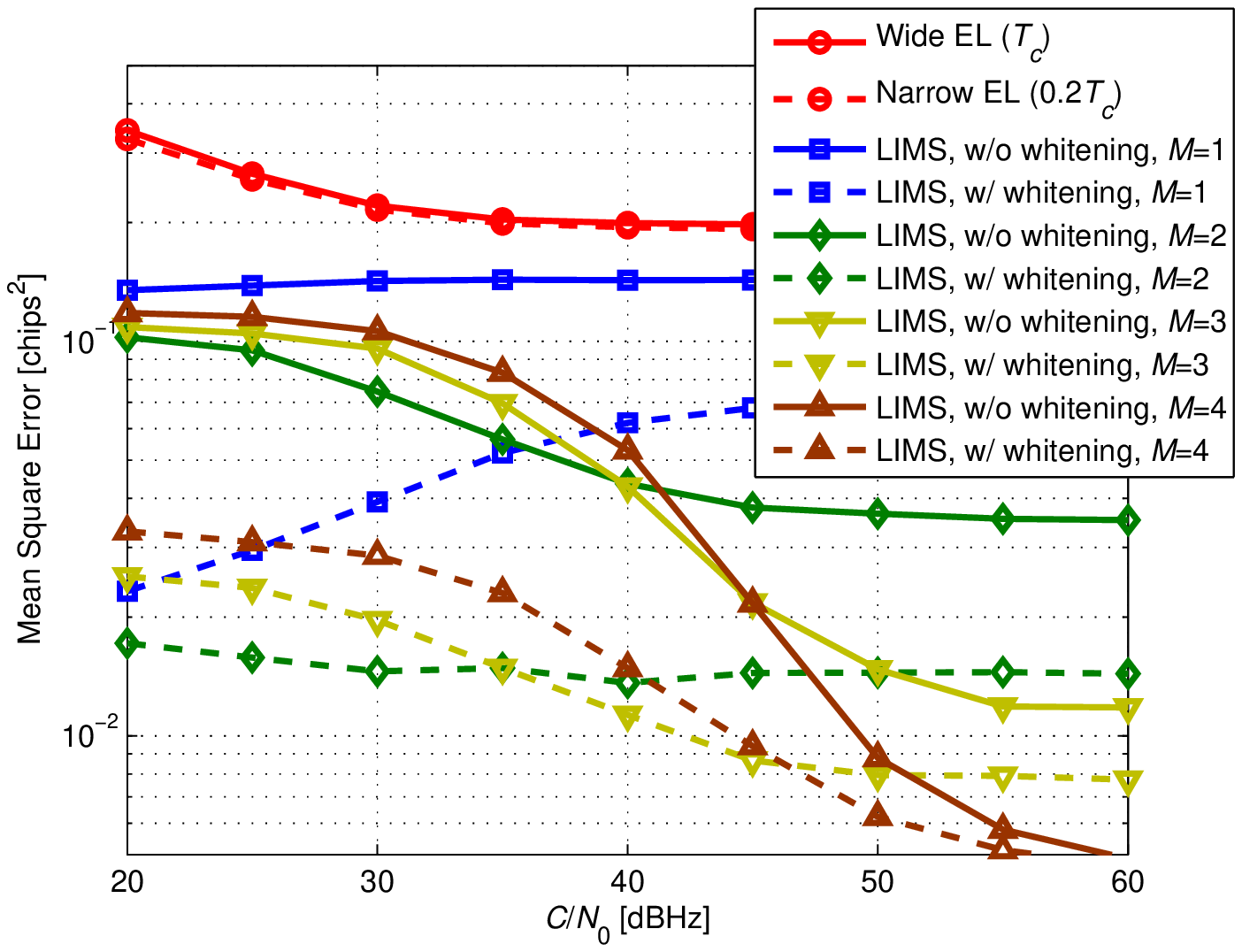,
width=8cm}\label{FIG:MultipathChInfB}} \caption{Mean square errors
of the first arrival path delay estimation: Multipath fading
channels without band limitation.} \label{FIG:MultipathChInf}
\end{center}
\end{figure}

The third test channel models considered for simulation are the
multipath fading channels. Two multipath fading channel models,
which are denoted by channels A and B, respectively, are
considered. Channels A and B have three and four paths,
respectively, and the power-delay profiles are given in Table
\ref{TAB:ChParam}.
%
%
Note that, for both channels, the average power
of the first arrival path is $7$dB lower than that of the strongest path.
With the given power-delay profiles, the channel coefficients are
generated independently as zero-mean circular symmetric complex
Gaussian random variables with corresponding variances for each
trial. The MSE performances of the EL discriminators and the LIMS
algorithms with different settings are plotted in Figs.
\ref{FIG:MultipathChInfA} and \ref{FIG:MultipathChInfB} for
channels A and B, respectively. In Figs. \ref{FIG:MultipathChInfA}
and \ref{FIG:MultipathChInfB}, it is observed that the EL
discriminators yield very poor performances. This is because the
EL discriminator tends to keep track of the largest magnitude point at
the correlator output, which would be located mostly near the
strongest path and hardly near the first arrival path. Due to the
same reason, the performance of the LIMS algorithm with $M=1$ is
also bounded. However, the LIMS algorithms with $M \geq 2$ show
much better performances since they can track multiple paths
simultaneously, and one of the tracked paths is possibly close to
the true first arrival path.


\begin{table} [b]
\begin{center}
\caption{Parameters of the three- and four-path fading channels.}
\label{TAB:ChParam}
\begin{tabular}{c|c|c|c|c}
\hline & \multicolumn{2}{c|}{Channel A} &
\multicolumn{2}{c}{Channel B} \\\cline{2-3}\cline{4-5} Path &
Relative
power (dB) & Delay ($T_c$) & Relative power (dB) & Delay ($T_c$)\\
\hline\hline

1&-7.0&0&-7.0&0\\
2&0&0.3&-7.0&0.2\\
3&-2.2&0.5&0&0.4\\
4&-&-&-2.2&0.6\\\hline
\end{tabular}
\end{center}
\end{table}

\psfull
\psdraft

\begin{figure} [t]
\begin{center}
\subfigure[Channel A]{\epsfig{file=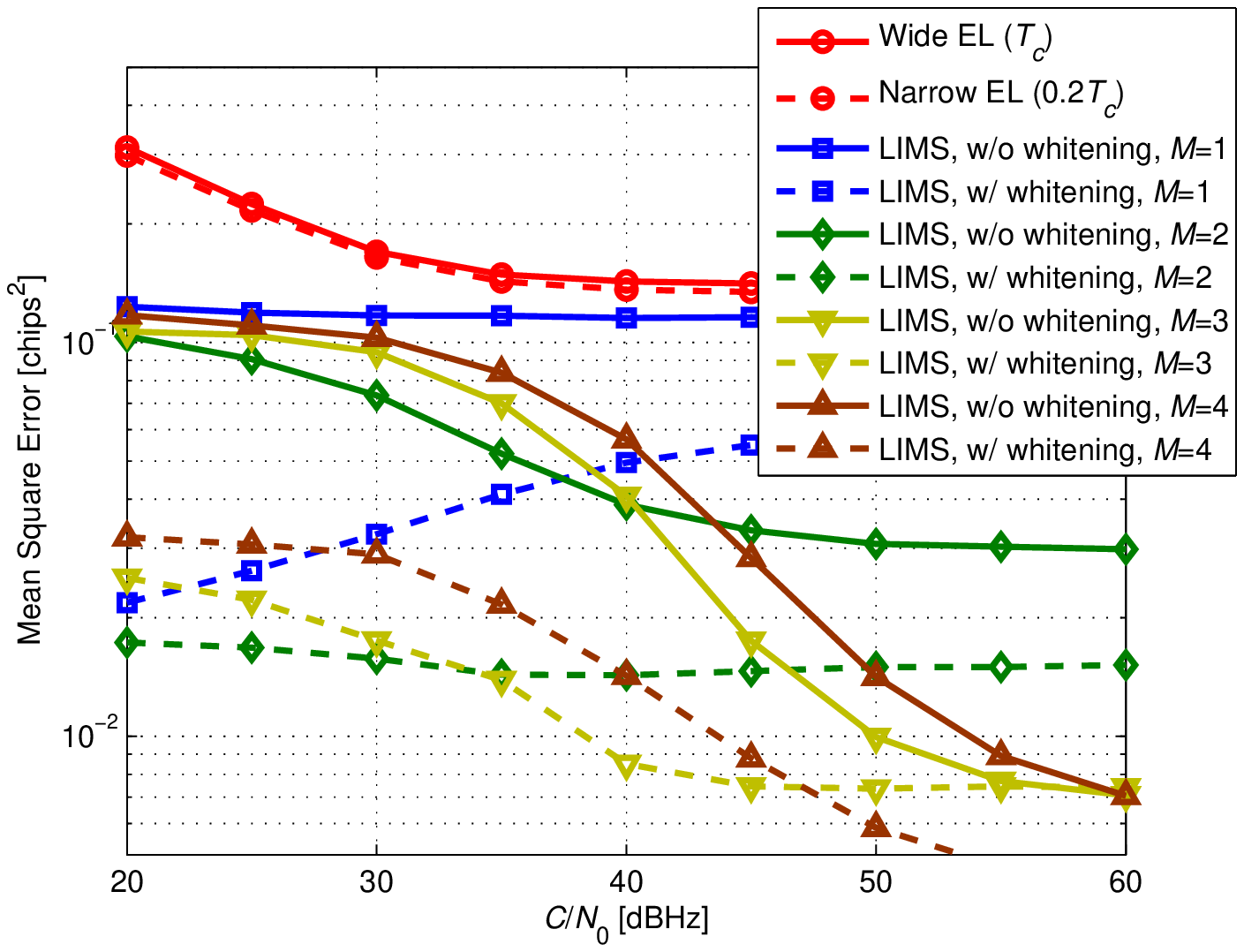,
width=8cm}\label{FIG:MultipathCh8A}} \subfigure[Channel
B]{\epsfig{file=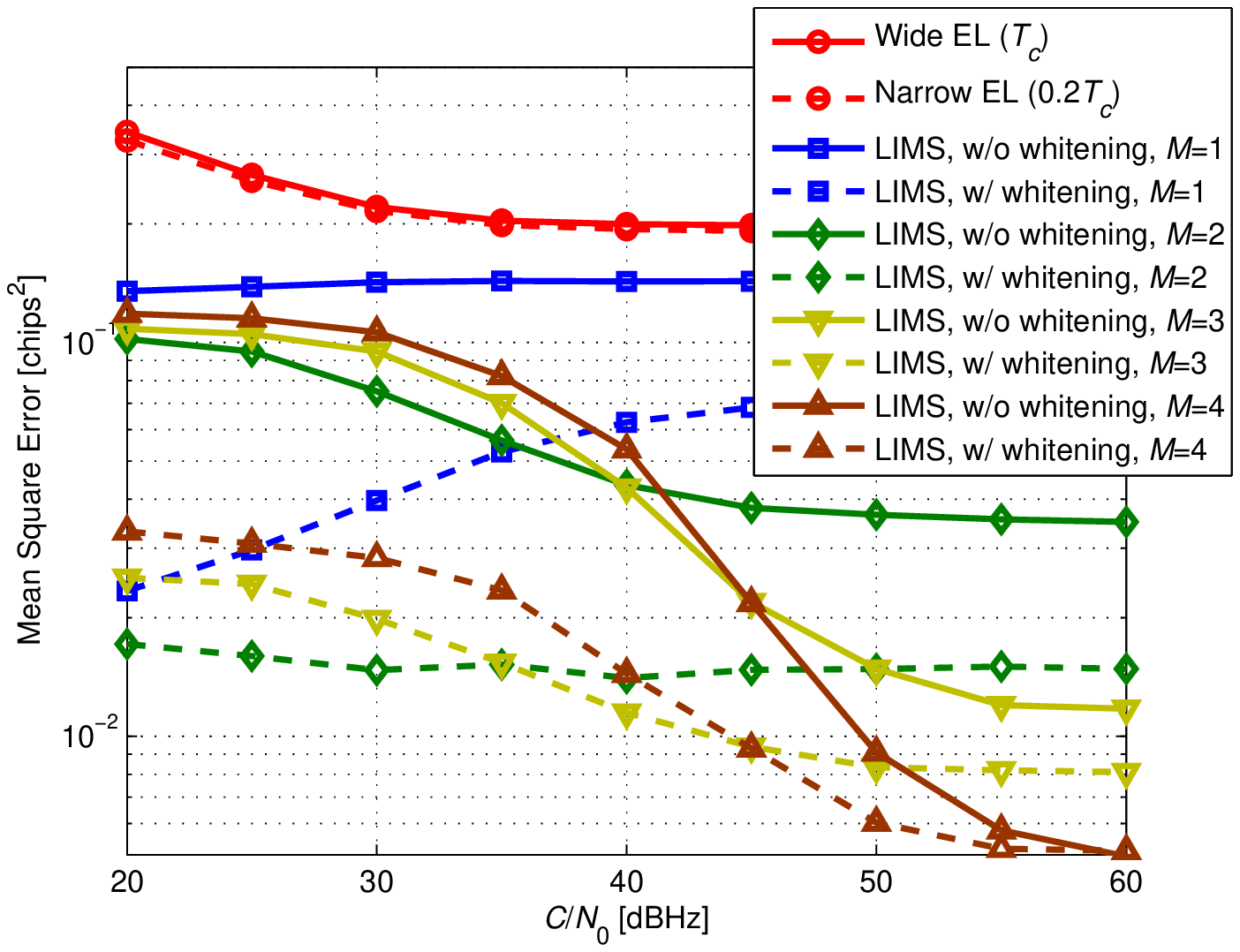,
width=8cm}\label{FIG:MultipathCh8B}} \caption{Mean square errors
of the first arrival path delay estimation: Multipath fading
channels with $8$MHz pre-correlation bandwidth.}
\label{FIG:MultipathCh8}
\end{center}
\end{figure}

\begin{figure} [t]
\begin{center}
\subfigure[Channel A]{\epsfig{file=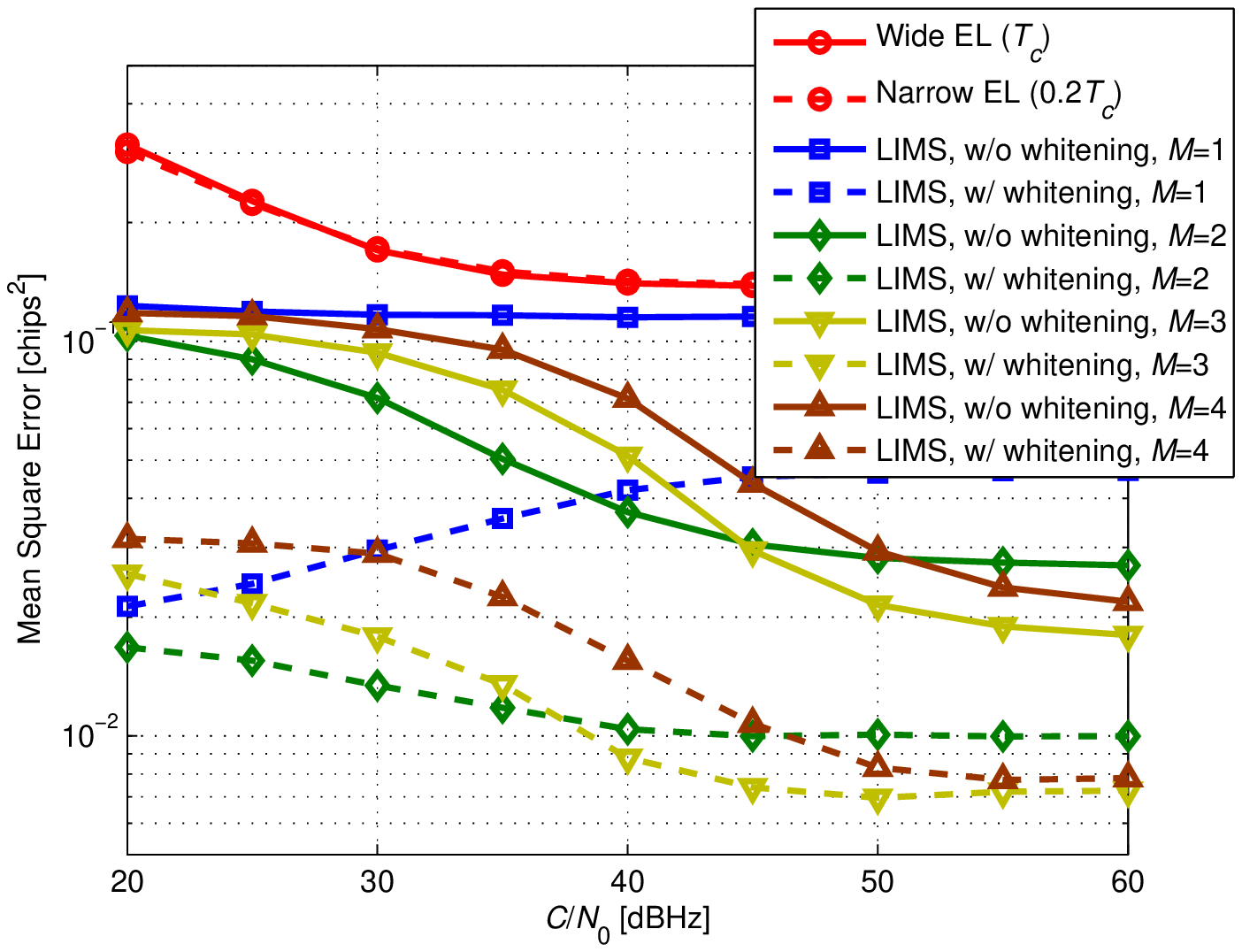,
width=8cm}\label{FIG:MultipathCh2A}} \subfigure[Channel
B]{\epsfig{file=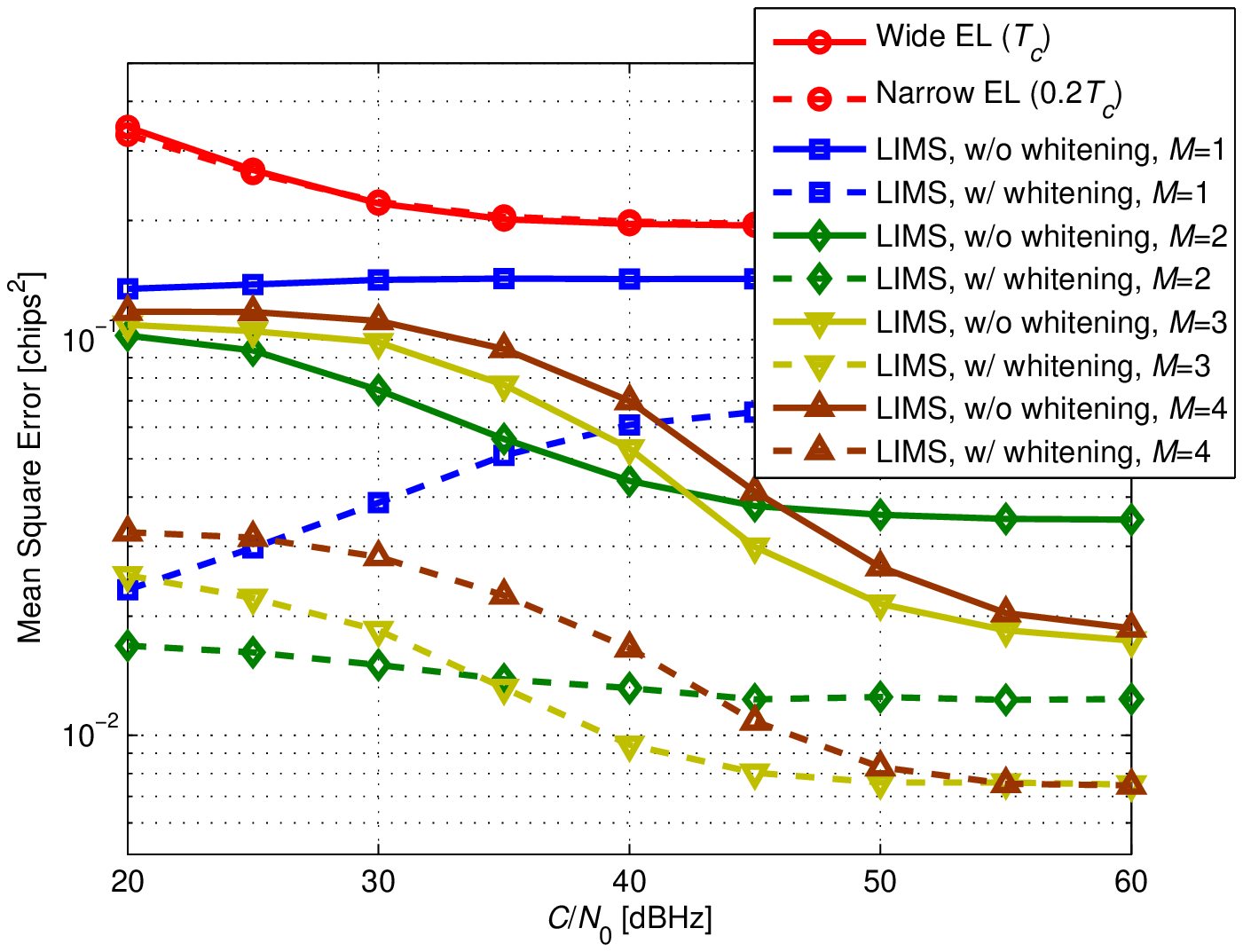,
width=8cm}\label{FIG:MultipathCh2B}} \caption{Mean square errors
of the first arrival path delay estimation: Multipath fading
channels with $2$MHz pre-correlation bandwidth.}
\label{FIG:MultipathCh2}
\end{center}
\end{figure}

The effect of limited bandwidth on the performance should also be
investigated. In practice, the received signal is bandlimited to
the pre-correlation bandwidth by a band pass filter before the
cross-correlation with the PN code sequence replica, and thus the
ACF may not have as sharp a triangular shape as
(\ref{EQ:AutoCorr}) has \cite[4.2]{YuText}. This causes a mismatch
between the assumed signal model and the true signal, and can
entail some performance loss. For the simulation, an ideal band
pass filter with a rectangular frequency response centered at the
carrier frequency of the signal is assumed. Figs.
\ref{FIG:MultipathCh8A} and \ref{FIG:MultipathCh8A} show the
results obtained for channels A
and B, respectively, when an $8$MHz pre-correlation bandwidth is applied. 
By comparing Figs. \ref{FIG:MultipathChInf} and
\ref{FIG:MultipathCh8}, it is found that there is no significant
loss when using an $8$MHz pre-correlation
bandwidth. 
The results with a $2$MHz pre-correlation bandwidth are also shown
in Figs. \ref{FIG:MultipathCh2A} and \ref{FIG:MultipathCh2B}.
There is apparent loss in this case, especially for the LIMS
algorithms with $M=3$ and $M=4$ at high $C/N_0$'s. However, the
performance loss is still not significant, and, therefore, one can
conclude that the LIMS algorithm is quite robust to the band
limitation.

\psfull
\psdraft

\psfull
\psdraft

%
%
%
%
%
%

\section{Conclusion} \label{SEC:Conclusion}

In this paper, the multipath channel estimation problem for direct
sequence spread spectrum signals was investigated. A new technique
based on the least-squares criterion, denoted as the least-squares
based iterative multipath super-resolution (LIMS) algorithm, was
proposed for the resolution of short delay multipaths. Unlike
conventional super-resolution techniques such as MUSIC and ESPRIT,
the LIMS algorithm does not rely on subspace decomposition, and,
therefore, has practical advantages. In particular, the LIMS
algorithm can work with a short observation of the received signal
regardless of the correlation between path coefficients. In
addition, due to its iterative operation, the LIMS algorithm is
suitable for recursive multipath tracking. We also presented a
view of the LIMS algorithm as a generalization of the conventional
EL discriminator for PN code phase tracking. Through numerical
simulations, it was shown that the performance of the LIMS
algorithm is superior to that of the EL discriminator in
estimating the delay of the first arrival path. As a result, the
LIMS algorithm can be a useful technique for applications such as
radio ranging, global navigation satellite systems (GNSSs), radio
propagation analysis, and time synchronization and multipath
tracking in wireless communication systems.

\appendix[The Cramer-Rao bound for the first arrival path delay
estimation] \label{APP:CRB}

From the signal model (\ref{EQ:SigModel}), the inverse Fisher
information matrix \cite[B.3.3]{StoicaText} for the estimation of
the channel coefficients and the path delays is computed as
\begin{equation}
{\bf F}^{-1} = \frac{T_i}{2N_0} \left( \Re \left\{ {\bf D}^H {\bf
C}^{-1} {\bf D} \right\} \right)^{-1} \text, \label{EQ:InvFisher}
\end{equation}
where ${\bf D} \in \mathbb{C}^{N \times 2K}$ is a concatenated
matrix defined as
\begin{equation}
{\bf D} = \left[ {\bf A} ({\bf t}_{\rm CH})\;\; {\bf B} ({\bf
c}_{\rm CH},{\bf t}_{\rm CH}) \right] \text,
\end{equation}
and ${\bf A} (\cdot)$ and ${\bf B} (\cdot,\cdot)$ are defined in
(\ref{EQ:Param}). The mean square error (MSE) of each parameter
estimation is lower bounded by the corresponding diagonal element
of the inverse Fisher information matrix. Thus, 
the MSE of the first arrival path delay estimation is lower
bounded by the $(K+1)$-th diagonal element of the inverse Fisher
information matrix (\ref{EQ:InvFisher}).

\end{document}